\documentclass[twocolumn,english,floatfix]{revtex4}
\usepackage[dvips]{graphicx}
\usepackage{hyperref}
\usepackage[T1]{fontenc}
\usepackage{amsmath}
\usepackage{amssymb}
\usepackage{babel}
\usepackage{latexsym}

\begin{document}
\title{Statefinder Revisited}
\author{ \'Emille E. O. Ishida }
%\email{emille@if.ufrj.br}
\affiliation{Universidade Federal do Rio de Janeiro, Instituto de
F\'{\i}sica, CEP 21941-972, Rio de Janeiro, RJ, Brazil}
%%% ----------------------------------------------------------------------

\begin{abstract}
The quality of supernova data will dramatically increase in the next
few years by new experiments that will add high-redshift supernova
to the currently known ones. In order to use this new data to
discriminate between different dark energy models, the statefinder
diagnostic was suggested \cite{S} and investigated by Alam \emph{et
al.}\cite{A} in the light of the proposed SuperNova Acceleration
Probe (SNAP) satellite. By making use of the same procedure
presented by these authors, we compare their analyzes with ours,
which shows a more realistic supernovae redshift distribution and do
not assume that the intercept is known. We also analyzed the
behavior of the statefinder pair \{\textit{r,s}\} and the
alternative pair \{\textit{s,q}\} in the presence of offset errors.
\end{abstract}
\maketitle
%%% ----------------------------------------------------------------------

\section*{Introduction}

Recent observations from type Ia supernovae measurements, cosmic
microwave background radiation and gravitational clustering suggest
the expansion of the universe is accelerated.

In order to explain this cosmic acceleration a form of
negative-pressure matter called \emph{dark energy} was suggested.
The simplest and most popular candidate is Einstein\textquoteright s
cosmological constant. Many others candidates for dark energy have
been proposed, including scalar fields with a time dependent
equation of state, quartessence, modified gravity, branes, etc.
Confrontation between these models and currently observational data
doesn\textquoteright t say much \cite{Q}, mainly because most of
them have  $\Lambda$CDM as a limiting case in the redshift range
already observed. The SNAP (SuperNovae Acceleration Probe) satellite
is expected to observe $\sim 2000$ supernovae per year with redshift
up to $z =1.7$. To differentiate models using the new available
data, Sahni \emph{et al.} \cite{S} introduced the \emph{statefinder}
diagnostic, that is based on the dimensionless parameters
\{\textit{r,s}\}, which are constructed with the scale factor and
its time derivatives.

In this work we applied the statefinder to a SNAP-like supernovae
distribution and analyzed its behavior in the presence of systematic
and random systematic, beyond statistical errors.

%%% ----------------------------------------------------------------------

\section{Dark Energy Models}
Assuming a Friedman-Robertson-Walker (FRW) metric, the
Einstein\textquoteright s equations reduce to:
\vspace{-0.5cm}

\begin{eqnarray}\label{einstein1}
    H^{2}&=&\frac{8\pi G}{3}\sum_{i}\rho_{i} - \frac{kc^{2}}{a^{2}}\\
    \frac{\ddot{a}}{a}&=& -\frac{4\pi G}{3}\sum_{i}(\rho_{i}+3\frac{p_{i}}{c^{2}})
\end{eqnarray}\*where \textbf{a} is the scale   factor  of  the  FRW  metric,
\textbf{H} is the   Hubble    parameter,    and  the  sum is  over
all the components present in the scenario in study.

In the following we assume that the  matter content of  the universe
is given by dark matter  ($p_{m}=0$) and  dark energy  with an
equation of state in the form $p_{x} = p_x (\rho_{x})$. We also take
$c=1$ and consider a  flat universe ($k=0$).

\begin{center}
\begin{table*}[!pt]
\caption{\label{tab:Table1} Redshift distribution. The value of z
represent the upper edge of each bin \cite{K}}
\begin{ruledtabular}
\begin{tabular}{cccccccccccccccccc}
z&0.1&0.2&0.3&0.4&0.5&0.6&0.7&0.8&0.9&1.0&1.1&1.2&1.3&1.4&1.5&1.6&1.7\\
N(z)&0&35&64&95&124&150&171&183&179&170&155&142&130&119&107&94&80\\
\end{tabular}
\end{ruledtabular}
\end{table*}
\end{center}

The  focus of our discussion will be in the four models listed
below:
\begin{enumerate}
  \item \textbf{Cosmological Constant} ($w_{x} = p_{x} /\rho_{x} = -1$)\\
  The   cosmological   constant   model    represents   a   constant    energy   density.
  In this model, the Hubble parameter has the form:
  \begin{equation}\label{modelo1}
    H(z)=H_{0}[\ \Omega_{m0} (1+z)^{3} + 1 - \Omega_{m0}]^{\frac{1}{2}}
  \end{equation}
  \item \textbf{Quiessence } ( $-1/3 > w_{x} = p_{x} /\rho_{x} = cte >-1$)\\
  This  is  the  next  simplest  example   of  dark  energy  model,
  and  gives  rise to a Hubble parameter like:
  \begin{equation}\label{modelo2}
    H(z)=H_{0}[\ \Omega_{m0} (1+z)^{3} + \Omega_{X0}(1+z)^{3(1+w)}]^{\frac{1}{2}}
  \end{equation}
  \item \textbf{Quintessence} ( w = w ( t ) )\\
  Representing   a   self-interacting   scalar   field   minimally
  coupled  to gravity. In this model, we have:
  \begin{eqnarray}\label{modelo3}
    \rho_{\phi}&=&\frac{1}{2} \dot{\phi}(z)^{2} + V(\phi(z))\\
    p_{\phi}&=&\frac{1}{2} \dot{\phi}(z)^{2} - V(\phi(z))
  \end{eqnarray}
  We shall focus on  a  special kind  of quintessence  model  that has  a  \emph{tracker}
  like solution,  with the following potential: $V(\phi)=\phi(z)^{-\alpha} (\alpha >0)$.
  In this case, it can be shown that the present energy density of the dark energy is
almost independent of initial conditions.
  \item \textbf{Chaplygin Gas}\\
  A different kind of solution is provided by the Chaplygin gas model.
  In this model, the Hubble parameter takes the form:

  \begin{equation}\label{modelo4b}
    H(z)=H_{0}\bigg[\Omega_{m0} (1+z)^{3} + \frac{\Omega_{mo}}{\kappa}\sqrt{\frac{A}{B}(1+z)^{6}}\bigg]^{\frac{1}{2}}
  \end{equation}
  where
  \begin{equation}\label{kappa}
  \kappa=\frac{\rho_{mo}}{\rho_{ch0}}
  \end{equation}

  The Chaplygin  gas  behaves  like a cosmological constant for small
  z (late times) and like pressureless dust for large z (early times).
\end{enumerate}

It is important to note that for  all  models  presented previously,
the luminosity distance is :

\begin{equation}\label{lumdist}
    \frac{D_{L}(z)}{1+z}=\int_{0}^{z}\frac{dz'}{H(z')}
\end{equation}\*with $H(z)$ given by the model dependent expressions presented
before.

%%% ----------------------------------------------------------------------

\section{The Statefinder}

The properties of dark energy, as we have seen, are very model
dependent. In order to differentiate between the presented models,
Sanhi \emph{et al.}\cite{S}, proposed the statefinder diagnostic.
The parameters, \{\textit{r,s}\}, are a complement to the already
known deceleration parameter, and help the discrimination when the
later contains degeneracies. By definition:
\begin{eqnarray}\label{q}
    q=-\frac{\ddot{a}}{aH^{2}}\equiv\frac{1}{2}(1+3w\Omega_{X})\\
\label{r}
    r\equiv\frac{\ddot{a}}{aH^{3}}=1+\frac{9w}{2}\Omega_{X}(1+w)-\frac{3}{2}\Omega_{X}\frac{\dot{w}}{H}\\
\label{s}
    s\equiv\frac{r-1}{3(q-\frac{1}{2})}=1+w-\frac{1}{3}\frac{\dot{w}}{wH}
\end{eqnarray}

%%% ----------------------------------------------------------------------

\section{Dark Energy from SNAP data}

Type Ia supernovae are considered standard candles, used to map the
expansion history, and its observations lead to the behavior of the
scale factor with time. In order to study the data in a model
independent way, we use a parametrization for the dark energy
density, presented by Sahni \emph{et al.} \cite{S}. We express the
energy density as a power series up to second order in z:
$\rho_{DE}=\rho_{c0}(A_{1} + A_{2}x + A_{3}x^{2})$, where $x = 1+z$.
For this Ansatz, the Hubble parameter takes the form:

\begin{equation}\label{ansatz}
    H(x)=H_{0}(\Omega_{m0}x^{3}+A_{1}+A_{2}x+A_{3}x^{2})^{\frac{1}{2}}
\end{equation}\*equation (\ref{ansatz}) together with equation (\ref{lumdist})
provides an expression for the luminosity distance, which we shall
investigate, using the statefinder parameters, in the light of a
SNAP-like experiment simulation.

To simulate the data, we use a binned approach for a SNAP
distribution shown in the Table I. We also include $300$ supernovae
in the first bin. These low redshift supernovae are expected from
the SNFactory (Nearby Supernovae Factory)  experiment and are
important in reducing the systematic errors. The SNFactory proposal
is to provide data to calibrate high redshift experiments, like the
SNAP, and then reduce the errors involved.

The luminosity distance of equation (\ref{lumdist}) is measured in
terms of the apparent magnitude, which can be written as:
\vspace{-0.3cm}
\begin{eqnarray}\label{eq:mag}
    m(z)&=&5\log{d(z)}+\nonumber\\
        & &[M+25-5\log{(H_{0}/(100km/s/Mpc)})]
\end{eqnarray}\*where M is the absolute magnitude of the supernova and the
expression in brackets is called \emph{intercept}.

Following what was done by Kim \emph{et al.} \cite{K}, we consider a
random irreducible systematic error of $0.04*(z_{med}/1.7)$ (here
{$z_{med}$ is the redshift in the middle of each bin),  added in
quadrature to a constant statistical error of $0.15mag$ and study
the behavior of the statefinder parameter with this synthetic data.
We performed a Monte Carlo simulation considering the intercept
exactly known and totally unknown. As a second step, we study the
situation where offset errors are present, and its consequences in
the statefinder.

\section{Results}

In the figures \ref{r-med-est} to \ref{q-med-offset} we present the
results from simulated data. According to SNAP's specifications, we
generated $500$ data sets having $\Lambda$CDM as a fiducial model.
For each of this experiments we calculated the best fitting
parameters $A_{1}$ and $A_{2}$ for equation (\ref{ansatz}) and
reconstructed the statefinders $r(z)$ e $s(z)$. The figures below
show the mean value of the parameters, which were calculated as:
\begin{eqnarray}\label{rmedio}
<(r)>&=&\frac{1}{500}\sum_{i=1}^{500}r_{i}(z)\\
<(s)>&=&\frac{1}{500}\sum_{i=1}^{500}s_{i}(z)\\
<(q)>&=&\frac{1}{500}\sum_{i=1}^{500}q_{i}(z)
\end{eqnarray}

\begin{figure}[!t]
\centering
 \includegraphics[width=6.5cm,
  height=4cm]{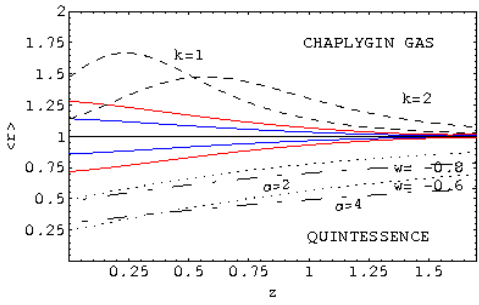}\\
  \caption{\footnotesize{Shows  $<r>$  as a function of redshift.
  The blue contour represent
$1\sigma$ and the red contour $2\sigma$ confidence levels, in the
presence of statistical errors with a known intercept.The line $<r>=
1$ represents the $\Lambda$CDM fiducial model. The dashed lines
above the $\Lambda$CDM are Chaplygin gas with parameters $k=1$ and
$k=2$. The dotted lines below it are quiessence models with $w=-0.6$
and $w=-0.8$, and the dotted-dashed lines are quintessence models
with $\alpha=2$ and $\alpha=4$.}}\label{r-med-est}
\end{figure}

\begin{figure}[!ht]
\centering
  \includegraphics[width=6.5cm,
  height=4cm]{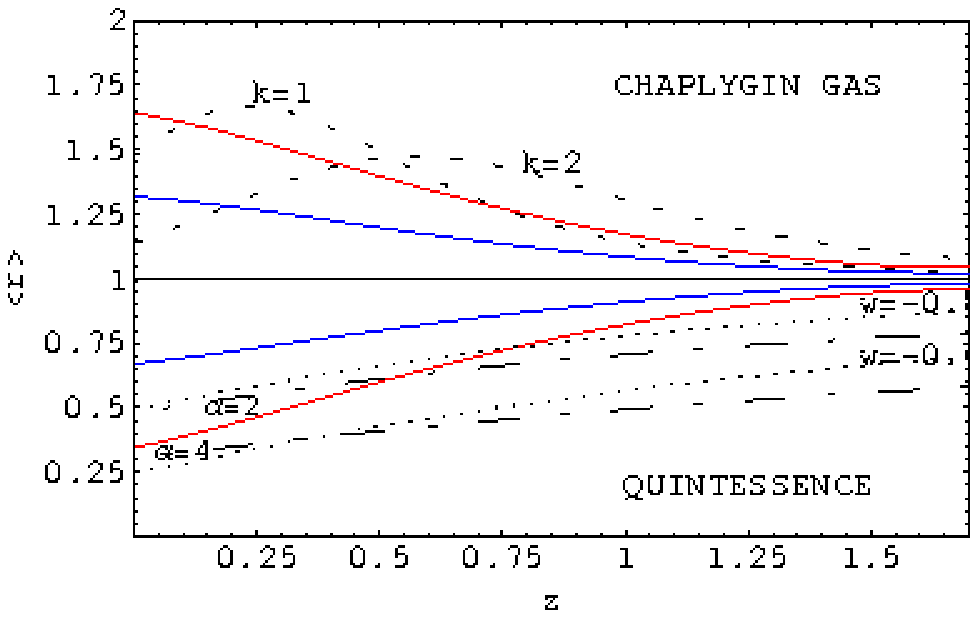}\\
  \caption{\footnotesize{Analog to figure \ref{r-med-est}, but here in the presence of random systematic
  and statistical errors, with an unknown intercept.}}\label{r-med-nk}
\end{figure}

\begin{figure}[!h]
\centering
  \includegraphics[width=6.5cm,
  height=4cm]{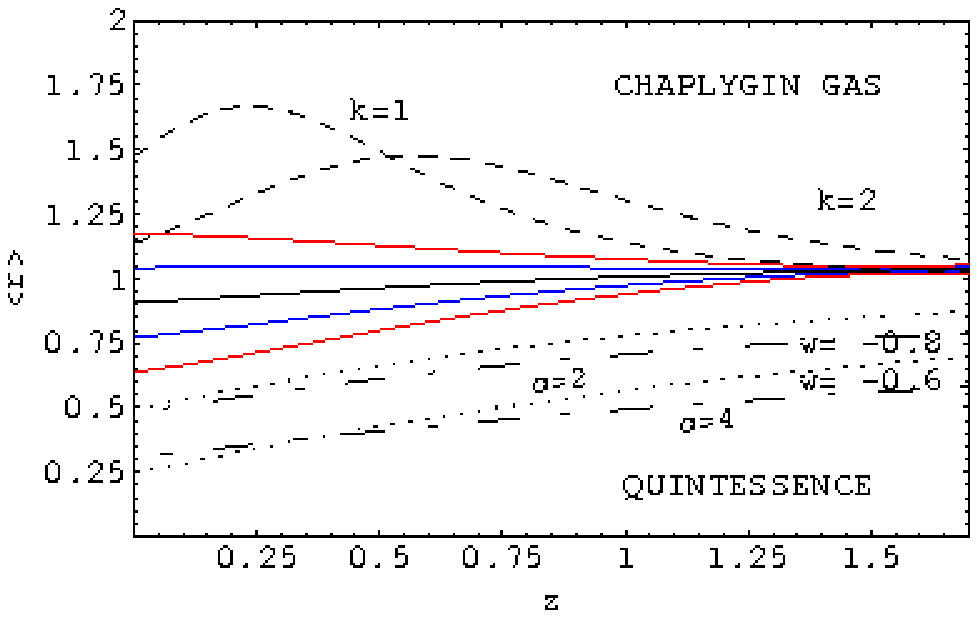}\\
  \caption{\footnotesize{Shows $< r >$ as a function of redshift. Again,
  the blue and red contours are $1\sigma$ and $2\sigma$ confidence levels,
  in these we considered a known intercept in the presence of statistical error
and a systematic error of $+0.03mag$.The dotted, dot-dashed and
dashed lines are the same as in figure
\ref{r-med-est}.}}\label{r-med-offset}
\end{figure}

\begin{figure}[!t]
\centering
  \includegraphics[width=6.5cm,
  height=4cm]{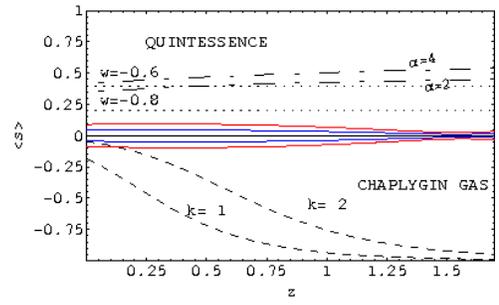}\\
  \caption{\footnotesize{Shows  $<s>$  as a function of redshift in the presence
  of statistical error. The colored contours, dotted,
  dot-dashed and full lines represent the same models as in figure \ref{r-med-est}.}}
\end{figure}\label{s-med-est}

\begin{figure}[!ht]
\centering
  \includegraphics[width=6.5cm,
  height=4cm]{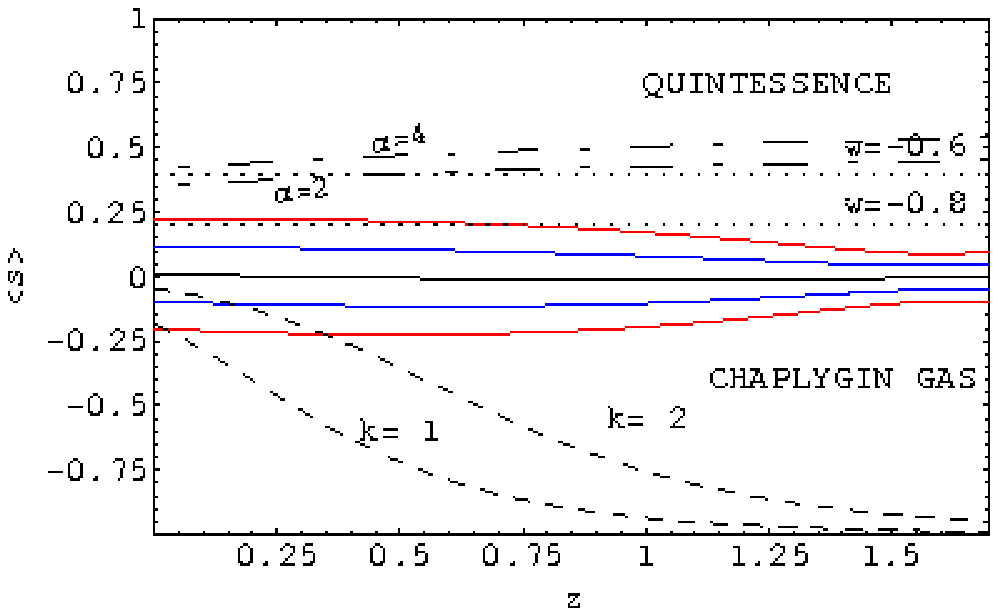}\\
  \caption{\footnotesize{Analog to figure \ref{s-med-est}, but here in the presence of random systematic
  and statistical errors, with an unknown intercept.}}\label{s-med-nk}
\end{figure}

\begin{figure}[!h]
\centering
  \includegraphics[width=6.5cm,
  height=4cm]{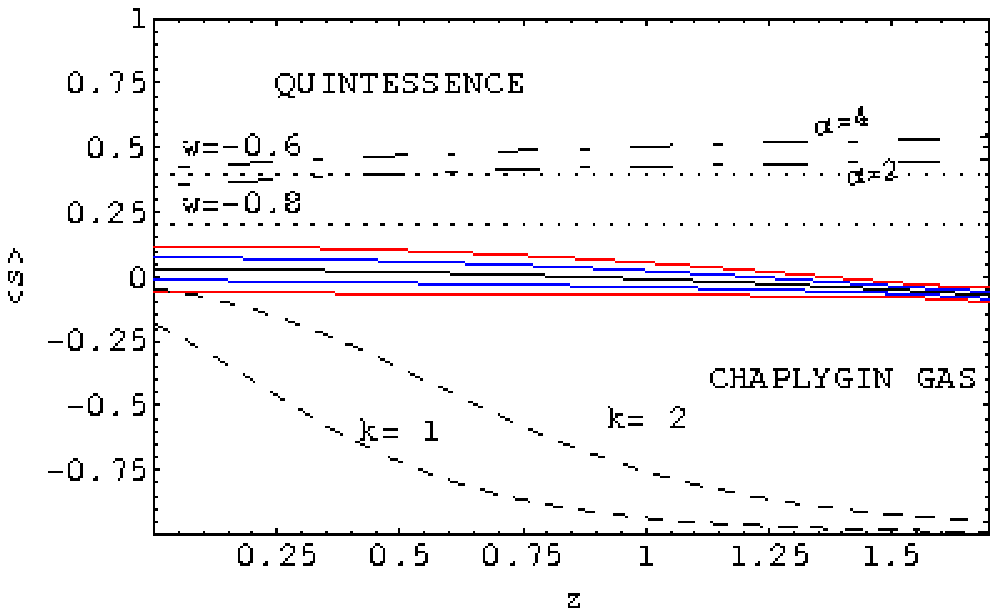}\\
  \caption{\footnotesize{Shows $< s >$ as a function of redshift. Again,
  the blue and red contours are $1\sigma $and $2\sigma$ confidence levels,
  in this we considered a known intercept in the presence of statistical error
and a systematical errors of $+0.03mag$.The dotted, dot-dashed and
dashed lines are the same as in figure
\ref{r-med-est}.}}\label{s-med-offset}
\end{figure}

\begin{figure}[!h]
\centering
  \includegraphics[width=6.5cm,
  height=4cm]{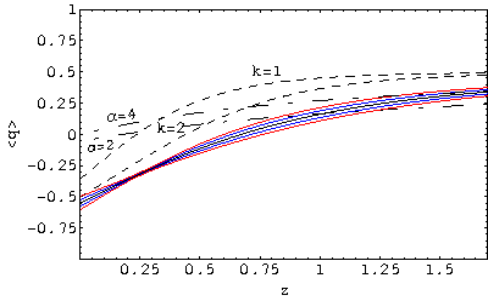}\\
  \caption{\footnotesize{Shows  $<q>$  as a function of redshift. The blue contour represent
$1\sigma$ and the red contour $2\sigma$ confidence levels. The
dashed lines are Chaplygin gas with parameters $k=1$ and $k=2$. The
dotted-dashed lines are quintessence models with $\alpha=2$ and
$\alpha=4$.}}\label{q-med-est}
\end{figure}

\begin{figure}[!ht]
\centering
  \includegraphics[width=6.5cm,
  height=4cm]{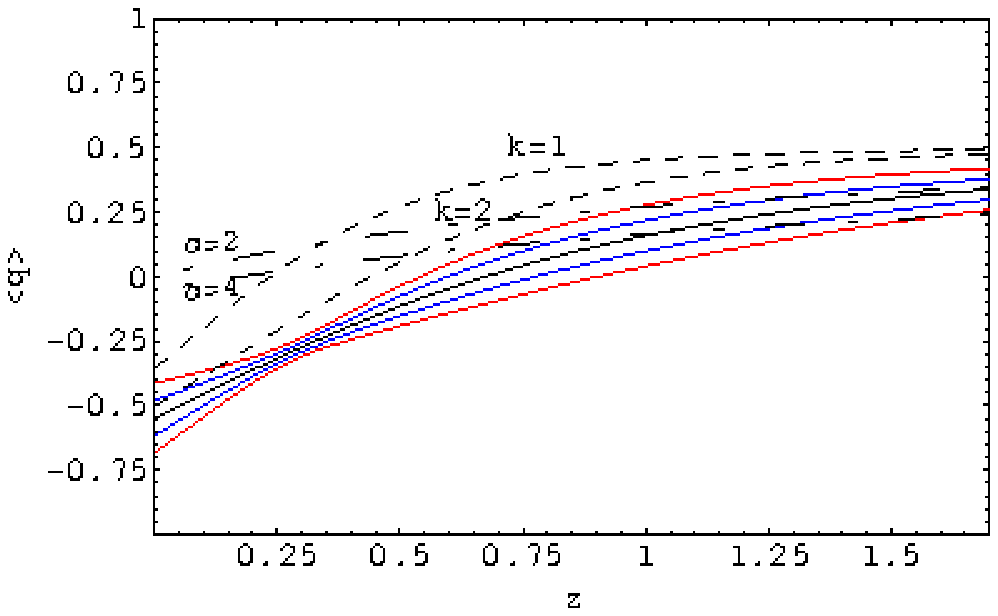}\\
  \caption{\footnotesize{Analog to figure \ref{q-med-est}, but here in the presence of random systematic
  and statistical errors, with an unknown intercept.}}\label{q-med-nk}
\end{figure}

\begin{figure}[!ht]
\centering
  \includegraphics[width=6.5cm,
  height=4cm]{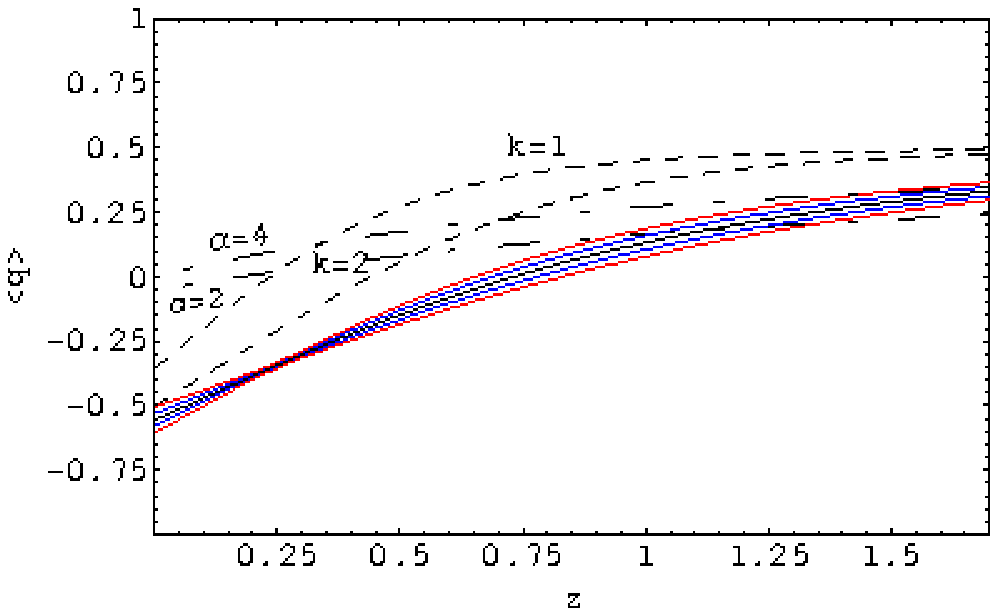}\\
  \caption{\footnotesize{Shows $< q >$ as a function of redshift. Again,
  the blue and red contours are $1\sigma$ and $2\sigma$ confidence levels,
  in these we considered a known intercept in the presence of statistical error
and a systematical error of $+0.03mag$.The dot-dashed and dashed
lines are the same as in figure
\ref{q-med-est}.}}\label{q-med-offset}
\end{figure}

Next, we investigated the integrated averaged of the cosmological
parameters, as suggested by Alam \emph{et al.} \cite{A}. For the
cosmological constant model, the parameters are constant, but for
many dark energy models the statefinder evolves (as it is clear in
the figures presented previously), the integration of this
quantities may then reduce the noise in the original data. The
integration was done as follows:

\begin{eqnarray}\label{rbarra}
\bar{r}&=&\frac{1}{z_{max}}\int_{0}^{z_{max}}r(z)dz\\
\bar{s}&=&\frac{1}{z_{max}}\int_{0}^{z_{max}}s(z)dz\\
\bar{q}&=&\frac{1}{z_{max}}\int_{0}^{z_{max}}q(z)dz
\end{eqnarray}
\\
where $z_{max} = 1.7$ . The expressions for \textit{r, s} and
\textit{q} were calculated using equations \ref{ansatz},  \ref{r}
and the first part of equation \ref{s} for each experiment. The
$500$ points obtained are plotted in figures \ref{rxs-est} to
\ref{sxq-offset}:

\begin{figure}[!ht]
\centering
  \includegraphics[width=6.5cm,
  height=4cm]{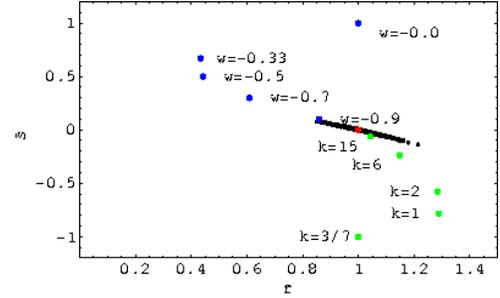}\\
  \caption{\footnotesize{Shows the variation of  $\bar{r}$ with  $\bar{s}$  , in the presence
of  statistical error when the intercept is known. The red dot is
the $\Lambda$CDM fiducial model. The blue dots above the
$\Lambda$CDM are quiessence models with
  $w = 0.0, -0.3,(-0.5), -0.7$ and $–0.9$ respectively,  from  top to bottom.
  The green dots below $\Lambda$CDM are  Chapligyn  gas  with  $k=(3/7), 1, 2, 6, 15$,
  from bottom to top.}}\label{rxs-est}
\end{figure}

\vspace{1cm}
\begin{figure}[!ht]
\centering
  \includegraphics[width=6.5cm,
  height=4cm]{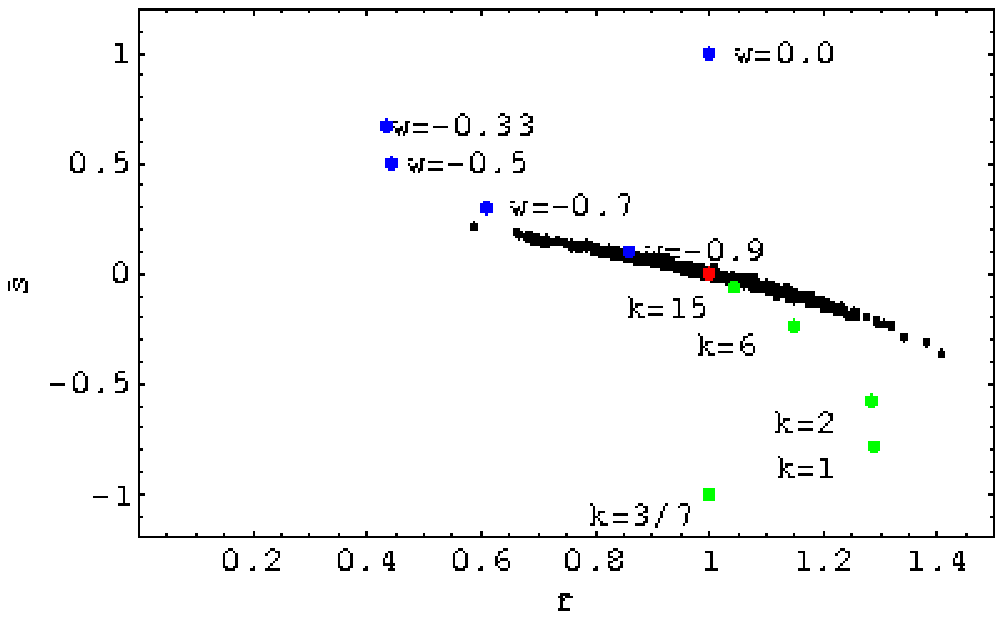}\\
  \caption{\footnotesize{This is analogous to figure \ref{rxs-est}, although here we
  considered an intercept not known and included statistical and random
  systematic errors.}}\label{rxs-nk}
\end{figure}

\vspace{1cm}
\begin{figure}[!t]
\centering
 \includegraphics[width=6.5cm,
  height=4cm]{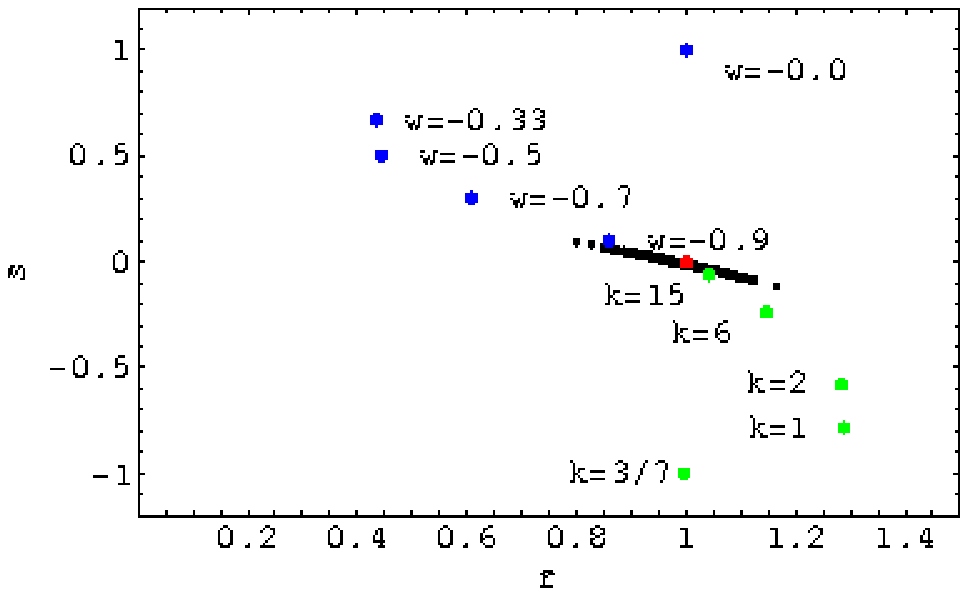}\\
  \caption{\footnotesize{Show the variation of  $\bar{r}$ with $\bar{s}$ when the intercept
  is known, in the presence of statistical error and systematic error of $+0.03mag$. The red,
   blue and green dots are the same as in figure \ref{rxs-est}.}}\label{rxs-offset}
\end{figure}

\begin{figure}[!ht]
\centering
 \includegraphics[width=6.5cm,
  height=4cm]{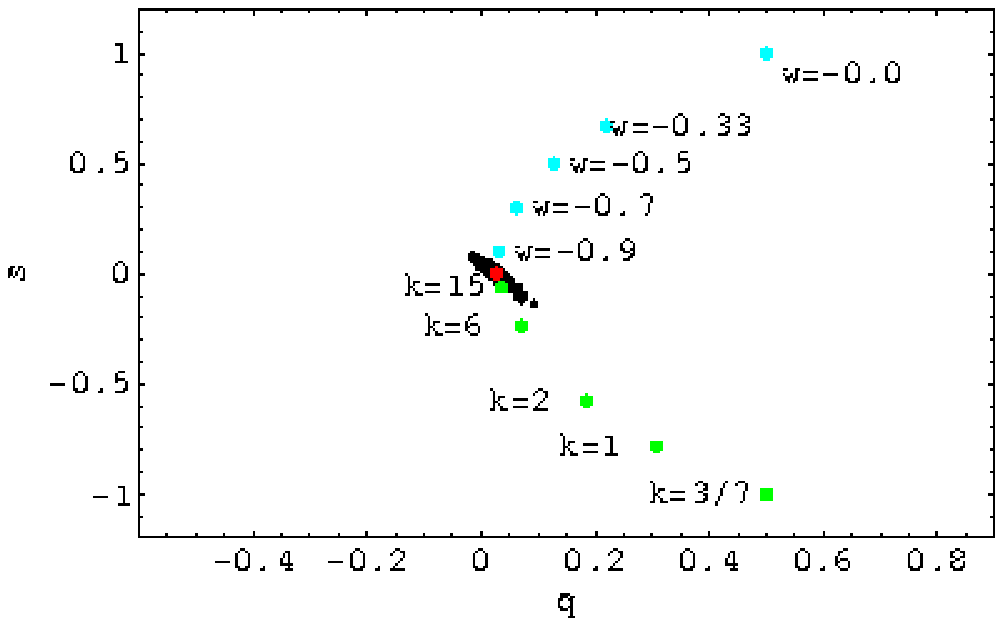}\\
\caption{\footnotesize{Shows the variation of  $\bar{s}$ with
$\bar{q}$, in the presence of  statistical error when the intercept
is known. The red dot is the $\Lambda$CDM
 fiducial model. The blue dots above the $\Lambda$CDM are quiessence models with
$w = 0.0, -0.3,(-0.5), -0.7$ and $–0.9$ respectively,  from  top to
bottom. The green dots below $\Lambda$CDM are  Chapligyn  gas  with
$k=(3/7), 1, 2, 6, 15$, from bottom to top.}}\label{sxq-est}
\end{figure}

\begin{figure}[!h]
\centering
 \includegraphics[ width=6.5cm,
  height=4cm]{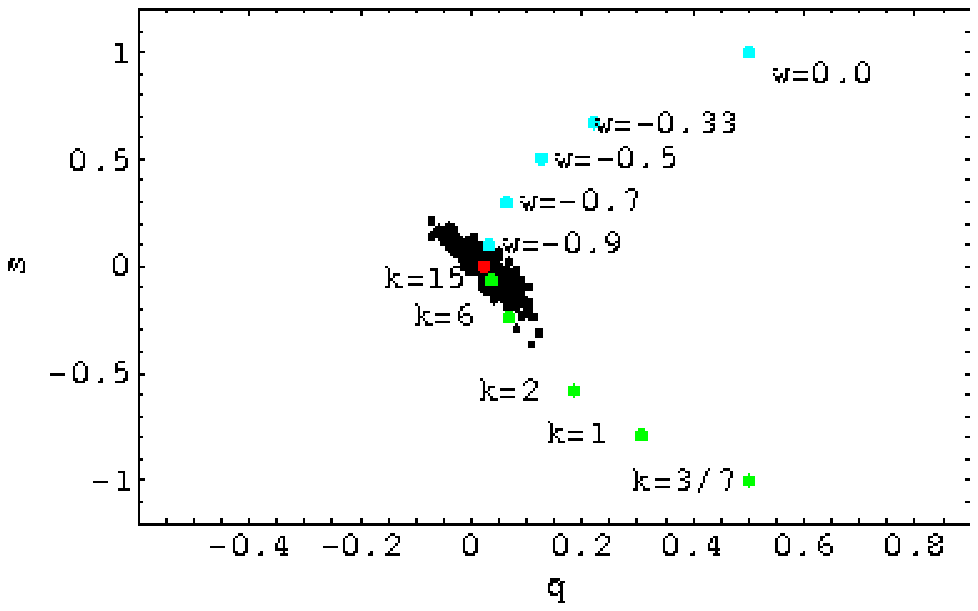}\\
  \caption{\footnotesize{This is analogous to figure \ref{sxq-est}, although here we
  considered an intercept not known and included statistical and random
  systematic errors.}}\label{sxq-nk}
\end{figure}

 \begin{figure}[!hhh]
 \centering
   \includegraphics[width=6.5cm,
   height=4cm]{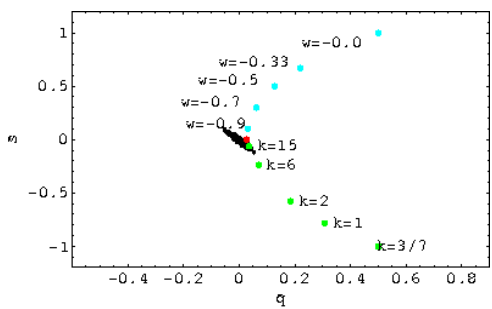}\\
   \caption{\footnotesize{Show the variation of  $\bar{s}$ with $\bar{q}$ when the intercept
   is known, in the presence of statistical and systematic error of $+0.03mag$. The red,
    blue and green dots are the same as in figure \ref{sxq-est}.}}\label{sxq-offset}
 \end{figure}

%%% ----------------------------------------------------------------------
\section{Discussion}

Our results suggest that the statefinder is a good diagnostic for
dark energy models, although some care must be taken in applying it
to data. As expected, the presence of random systematic error and an
unknown intercept just added more possible models than those allowed
by statistical errors only. There is little problem in this, once
the fiducial model is at least within the $2\sigma$ contours
(Figures \ref{r-med-nk}, \ref{s-med-nk} and \ref{q-med-nk}).

The presence of offset errors had different outcomes. For a positive
offset, the parameters \textit{r}  and \textit{s} suffered a small
reduction in relation to the fiducial model, but in different
redshift ranges (r for small and s for high redshift). A consequence
of this arises when we compare the integrated averages of the pair
\{\textit{r,s}\}. In Figure \ref{rxs-offset} the points are shifted
to the negative direction of both axes of the  ellipse, resulting a
data set where the fiducial model (red point) is on the edge of the
distribution. The same kind of shift is observed for a negative
offset, but in this case to the positive direction of the axes. So,
in order to use the statefinder as a diagnostic, we must be able to
control the offset error below $0.03mag$.

It is interesting to observe the behavior of the deceleration
parameter when systematic errors were involved. It does present a
very small reduction (positive offset) at high redshift, but it is
irrelevant in front of that suffered by \emph{r} or \emph{s}.
Comparing figures \ref{rxs-est} and \ref{sxq-est} we could say, as
suggested by Alam \emph{et al.} \cite{A}, that the pair
\{\textit{q,s}\} is even a better diagnostic than \{\textit{r,s}\},
once it restricts the area of the phase space filled by the data.
However, if there is a systematic error present, the points will be
shifted to the negative direction of the \emph{q} axis only (figure
\ref{sxq-offset}), letting the distance between the data and the
fiducial model bigger than those in figure \ref{rxs-offset}.

Therefore, we concluded that the statefinder pair \{\textit{r,s}\}
is a better diagnostic than the pair \{\textit{s,q}\}, when the
involved systematic errors are not random.

%%% ----------------------------------------------------------------------
\section*{Acknowlegments}

The author is grateful to Ioav Waga for suggesting this problem and
for extensive discussion and encouragement throughout the course of
the work. This work was supported by the Brazilian research agency
CNPq.

%%%%-------------------------------------------------------------------------

\end{document}